\pdfoutput=1
\documentclass[onecolumn,12pt, draftclsnofoot]{IEEEtran}

\usepackage{amsmath,amssymb,amsfonts,mathrsfs,bm}
\usepackage{amstext}
\usepackage{upgreek}
\usepackage{multicol}
\usepackage{indentfirst}
\usepackage{graphicx}
\usepackage{paralist}
\usepackage{multirow}
\usepackage{tabularx}

\usepackage{cite}
\usepackage{booktabs}
\usepackage{subfigure}

\IEEEoverridecommandlockouts

\begin{document}
\title{Caching at the Wireless Edge: Design Aspects, Challenges and Future Directions}
\author{Dong Liu$^{\dag}$, Binqiang Chen$^{\dag}$, Chenyang Yang$^\dag$, and Andreas F.
Molisch$^\ddag$
\thanks{$^\dag$School of Electronics and Information Engineering, Beihang University, Beijing 100191, P. R. China. Emails: \{dliu,chenbq,cyyang\}@buaa.edu.cn}
\thanks{$^\ddag$Ming Hsieh Department of Electrical Engineering, University of Southern California, Los Angeles, CA 90089-2565, USA. Email: molisch@usc.edu}
\thanks{This work was supported in part by the National Natural Science Foundation of China (NSFC) under Grant 61120106002 and National Basic Research Program of China (973 Program) under Grant 2012CB316003. The work of D. Liu was supported by the Academic Excellence Foundation of BUAA for PhD Students. The work of A. F. Molisch was supported by the National Science Foundation.} 
}
\maketitle
\begin{abstract}

Caching at the wireless edge is a promising way of boosting spectral efficiency and reducing energy consumption of wireless systems. These improvements are rooted in the fact that popular contents are reused, asynchronously, by many users. In this article, we first introduce methods to predict the popularity distributions and user
preferences, and the impact of erroneous information. We then discuss the two aspects of caching systems, namely content placement and delivery. We expound the key differences between wired and wireless caching, and outline the differences in the system arising from where the caching takes place, e.g., at base stations, or on the wireless devices themselves. Special attention is paid to the essential limitations in wireless caching, and possible tradeoffs between spectral efficiency, energy efficiency and cache size.

\end{abstract}

\section{Introduction}

The main trend of improving spectral efficiency (SE) for
5th generation (5G) cellular networks is an enhanced exploitation of spatial
resources.
By reducing the distance between base stations (BSs) and
users, order-of-magnitude increase in network throughput can be expected. However, the effectiveness of this approach
relies on high-speed backhaul connectivity to every single BS, whose capacity must exceed the aggregated data
rate of all its served users, which often is not practical for the reason of cost.

An alternative approach to improving SE is to reduce the unnecessary traffic load by rethinking the characteristics of the traffic itself and coming back to the most basic questions. Should all the data be conveyed right after they are generated or even requested?

The major driver of the exponential traffic growth in today's wireless
networks is mobile Internet, which inherits a large portion of traffic from the wired Internet. The majority of such traffic is content delivery, which is non-real-time in nature. In fact, it has been reported that video on demand (VoD), a major class of content dissemination traffic, will generate more than 69\% of
mobile data traffic by the end of 2019. Different from conventional  communications, the content generation of this traffic is usually decoupled from the content delivery, both in terms of source-destination and the time instance of generation-request.

Another key feature of the content delivery traffic is that a few
popular contents account for most of the traffic load, and are requested by many users at different times. With predicted content popularity endowed by
big data analytics, it is advantageous to cache popular contents \emph{locally} before the requests truly arrive, directly at the wireless edge, e.g., at BSs or even end user
devices~\cite{Niki13,Procach14}. This is all the more attractive as trend on the rapid growth of storage capacity of devices enables such caching at a relatively low cost.

Caching in the wired Internet is well established and has shown to reduce  latency and energy consumption. Similar benefits can be achieved by caching at the wireless edge, which can improve both SE and energy efficiency (EE) by precaching at users or further enabling device-to-device (D2D) communications, and by caching at small BSs (SBSs) to eliminate the backhaul bottleneck, possibly with a higher gain. This is because content is available {\em locally}, instead of requiring redundant traverse across backhaul links (not to mention transport across the Internet and mobile core network). However, the limitations of wireless networks need to be taken into account.

While the potential of  local caching has been revealed by several recent investigations ~\cite{Niki13,Ali13,Procach14,JMY15,Dong}, the current article provides an overview of the state-of-the-art challenges and possible solutions of caching at the wireless edge, which differs from~\cite{wang2014cache} that focused on caching at mobile core networks. To identify what is unique in wireless caching, we address the similarity and emphasize the difference between local caching in wired and wireless edge.

For both wired and wireless networks, caching consists of two closely-coupled problems: {\em content placement} and {\em content delivery}. The content placement problem includes determining the size and location of each cache, selecting which content from a library to place at which nodes, and how to download the content to these cache nodes. The content delivery problem is how to convey a content to a user that requests it.

Since local caching aims to improve the quality of experience
(QoE) of users or improve network performance without
compromising QoE, the benefits and the solutions of caching highly depend on the traffic characteristics, e.g., the users' demand profile and quality requirement. Therefore, we first introduce \emph{content popularity} and \emph{user preference}, and discuss the QoE metrics of two representative types of content delivery traffic and the resulting challenge.
Then, we address the differences in content placement and content delivery in wireless and wired networks. Next, we compare the SE and EE gains from caching at wireless and wired edge, and illustrate the fundamental tradeoffs and design aspects of caching at the wireless edge by simulation for two example systems. Finally, concluding remarks are provided.

\section{Key Features of Content Delivery Traffic}

\subsection{Content Popularity and User Preference}
Using big data analytics, the statistical patterns of content requests, both in aggregate form, and on a per-user basis, can be predicted and play a key role in the design of caching.

\subsubsection{Content popularity}

By popularity we mean the ratio of the number of requests for a
particular content to the total number of requests from users -- usually obtained for a certain region  during a given period of
time. It has been reported that the popularity of content follows a Zipf
distribution, a sort of power law distribution~\cite{Tatar14}, which can be characterized by the content
catalog size $N_f$ and a skewness parameter $\beta$. In general, the content popularity distribution
changes at a much slower speed than the traffic variation of cellular networks.
Consequently, it is usually approximated as constant over long
time (e.g., one week for movies, and two or three hours for news~\cite{Niki13}). Another important fact is that global popularity in a large region (say in a city or even a country) is often different from local popularity in a small region (say in a campus)~\cite{Dey14}.

Popularity prediction is an active research field recently, as it is beneficial to many applications such as network dimensioning and on-line marketing. Many prediction methods have been proposed, e.g., cumulative
views statistics based on the popularity correlation over time
\cite{Tatar14}. A special difficulty in wireless networks is the fine spatial granularity at which content popularity often needs to be known. Specifically,
predicting the content popularity in the coverage of a BS is challenging because the
users associated with a BS is dynamic and the number of cumulative requests is
limited during the lifetime of popular content.

\subsubsection{User preference}

The user preference profile comprises the probability that each content is requested by a specific user
during a certain period, and differs among  individuals. This comes from the
fact that a user usually has strong preference toward specific content categories~\cite{Dey14}.

The preference of a user can be predicted by machine learning (say collaborative filtering) based on the
historical content requests of the user and the similarity among users~\cite{shi2014}. This has been extensively studied  in recommendation systems and is a hot topic nowadays for more general applications.

\subsubsection{Prediction uncertainty}

The prediction accuracy of the content popularity and user preference affects the performance of proactive
caching. Specifically, the erroneous information reduces the probability of finding the requested files in the cache, called \emph{cache-hit probability}, a metric usually used to reflect caching performance. As a result, it reduces the performance gain from caching and introduces extra cost, e.g., the energy consumed at backhaul and BS for improper
content placement.


\subsection{VoD and File Downloading}
Content delivery traffic includes file downloading (e.g., software or data library update, music or video download) and video streaming.
For file downloading, the content is consumed after the complete file has been delivered, where download time is often used as a metric to reflect the QoE. For video streaming, the user starts to play the video immediately after sending the request,
where low initial delay, requested video quality, and few stalls during playback should be guaranteed.

Dynamic adaptive streaming over HTTP is common to reduce stalling under varying network congestion by providing various quality levels of videos in content server.
Caching at the wireless edge to support video
streaming service is more challenging.
Multiple versions of a video with different quality levels need to be cached by either storing differently compressed versions of the same video, or by using difference encoding such that a better quality video can be reconstructed from the lower quality video plus an enhancement layer.
Moreover, a mobile user may retrieve only a partial segment of a video from a BS  according to the video playback process and the moving speed. With limited
cache size, caching each complete video at every BS allows only a smaller number of distinct videos to be stored, while caching partial video may
cause the cached part to be useless for catching up with the playback process.

\section{Content Placement: Difference Between Caching in Wireless and Wired Edge}
In this section, we address the possible benefits, tradeoffs, and research issues of content placement at the wireless edge, by highlighting the difference with that at the wired edge, which comes from the architecture of wireless networks, nature of wireless channels, and user mobility.

\subsection{Caching at BSs or Users: Benefits and Tradeoffs}
\begin{figure}[htb]
	\centering
	\includegraphics[width=0.75\textwidth]{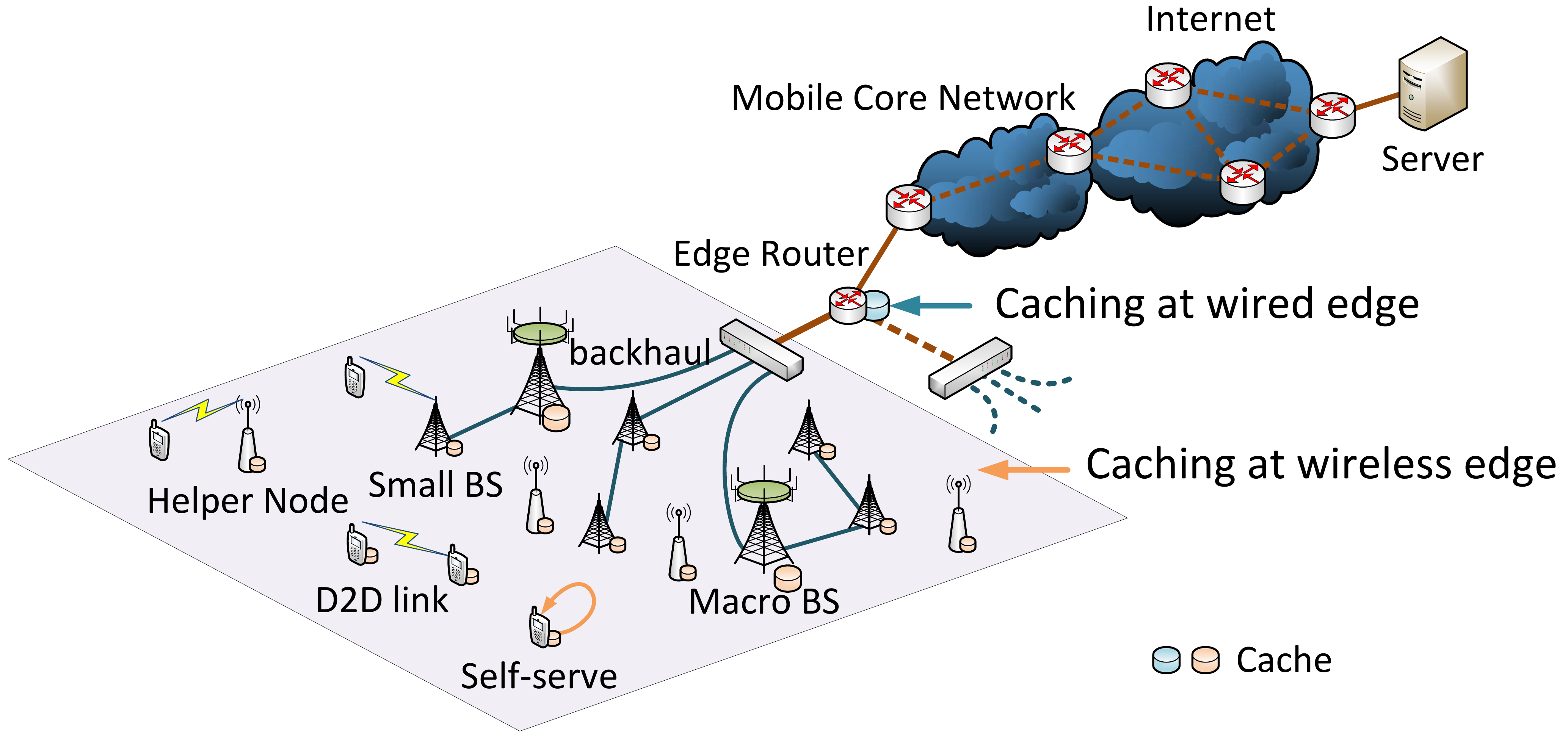}
	\caption{Local caching and content delivery at wireless edge.}
	\label{fig:archi} 
\end{figure}

In wireless networks, caches can be installed in macro BSs (MBSs), SBSs (say, pico or femto BSs), relays, or at users devices, see Fig. \ref{fig:archi}.

\subsubsection{Caching at BSs}
Compared with caching at evolved packet core (EPC) or at higher level, caching at existing MBSs and SBSs essentially plays the role of replacing backhaul links, and hence alleviates the backhaul congestion. Moreover, a new type of SBSs without any backhaul connections, called \emph{helpers}~\cite{Niki13}, enable flexible and cost-effective deployment to deliver popular contents.

Because increasing cache size can increase the cache-hit probability,
and hence lower the required backhaul capacity, there is a tradeoff between cache
size and backhaul capacity.
When the backhaul capacity is a bottleneck, increasing cache size is able to increase throughput. This translates to a tradeoff between the cache size and the network SE.

Since caching at the BSs can reduce the traffic in backhaul,
mobile core network and wired Internet, it can also improve the EE of the network, if power efficient cache hardware such as high-speed solid state disk is used.

%

As an important difference to wired networks, the performance gain of caching at the wireless edge highly depends on the inter-cell interference (ICI) level. In one example, the maximal EE gain of caching over not caching will be 575\% if the ICI can be completely removed and 250\%
without any ICI coordination~\cite{Dong}. When further considering overlapped coverage of densely deployed SBSs or heterogeneous networks and user mobility, the performance of caching in the wireless edge are yet to be fully exploited.

\subsubsection{Caching at users}
Precaching contents at user terminals such as smartphones, tablets and laptops has long been applied as a technique to improve QoE~\cite{Higgins12}. Recently, it is also proposed to offload wireless traffic.

With known content popularity, a BS can push the popular contents to all users via broadcast~\cite{LHui14}. With known user preference, the BS can pre-download favorite contents to some users via unicast.  According to whether the
contents are placed to a user via unicast~\cite{Higgins12} or broadcast~\cite{LHui14}, in this article we classify \emph{precaching} into {\em prefetching} and {\em pushing}.

When a user requests a content already cached at its local storage, the content can be retrieved from the cache (referred as
self-serve as in Fig. \ref{fig:archi}), with zero delay and without generating interference to other users.
Otherwise, the content can be conveyed to the user via unicast. In this way, precaching can improve QoE of all users either directly or indirectly, and improve wireless throughput by offloading.

In practice, a user may not be willing to contribute a large portion of its storage space for caching files. Then, pushing according to content popularity in a cell may not yield high cache hit probability, $p_h$. For instance, when the content popularity is Zipf distributed with  $N_f=1000$ and $\beta$ = 0.8, and 10 files can be pushed to a user, $p_h = ({\sum_{i=1}^{10} i^{-0.8}})/({\sum_{i=1}^{N_f}i^{-0.8}}) = 23\%$.
How to motivate users to cache more contents deserves investigation.

By further exploiting D2D communications, some users can share their cached contents to help improve the QoE of other users in  proximity and offload the
traffic~\cite{Niki13}.
If the requested
file is not cached at the local storage but cached at a nearby user, a D2D link can be established to deliver the content.

Recent results have demonstrated pronounced performance gain brought by cache-enabled D2D communications. Under many parameter settings, cache-enabled D2D provides a network throughput that increases linearly with the number of users (or, equivalently, a per-user throughput independent of the number of users)~\cite{JMY15}. This makes this technology one of the most promising methods for achieving order-of-magnitude improvements in SE. Besides, a tradeoff between throughput and outage is possible in cache-enabled D2D networks~\cite{JMY15}.

However, a high offloading ratio comes at the cost of the energy consumed at users, especially those acting as D2D transmitters. This may lead
to a tradeoff between the offloading ratio and  the energy consumption.

\subsection{Caching Policy in Wireless Edge: Unique Features}
Similar to caching in the wired networks, appropriate caching policy in wireless networks is critical for achieving the potential gain.

Caching policy can be either \emph{reactive} or \emph{proactive}. A \emph{reactive} caching
policy determines whether to cache a particular content after it has been requested according to certain replacement algorithms~\cite{Dey14}. A \emph{proactive} caching policy determines
which contents should be cached at each node before they are requested  based on the predicted users 
demand profile~\cite{Niki13,Procach14,Dey14}.
For proactive caching, the decision to cache among multiple nodes can be jointly optimized, and hence
the caching gain is high with perfect prediction. With prediction errors, however, the cache-hit probability will reduce, and
proactive caching may perform worse than reactive caching ~\cite{Gharaibeh2015Provably}.

\subsubsection{Caching at BSs}
Because the coverage of a BS is much smaller than an EPC or server and the connectivity between BS and user is with high uncertainty, caching policy design at BSs is more challenging.

\emph{Low cache-hit probability:} The size of cache and the number of requests at each BS are much smaller than those in the core network or Internet. As a result, the reactive caching policies designed for Internet are not effective for caching at the BSs~\cite{Dey14}. Designing proactive caching policy for each BS independently, e.g., each BS caching the most popular contents, may result in insufficient utilization of caches. 

One way to cope with this problem is to enable BSs to share the cached contents through a backhaul link, i.e., \emph{cooperative caching}~\cite{wang2014cache,Gharaibeh2015Provably}. If the requested content is not in the cache of the local BS, by retrieving the requested content from the caches of adjacent BSs instead of from the server, the delivery cost and latency can be reduced, and the overall cache-hit probability can be improved. Such an approach is more likely to be viable for MBSs that have high-capacity fiber connection to share data, but is hard to implement for SBSs.

Due to the openness of wireless channels, the coverage of the SBSs are often overlapped. This indicates that a user is able to fetch contents from multiple caches and hence the equivalent cache size seen from the user is increased. Based on this observation, caching policies for adjacent BSs can be jointly optimized to increase the cache-hit probability without data sharing over backhaul links, which is referred to as \emph{distributed caching}~\cite{Niki13}.

\emph{Topology uncertainty:} Different from the wired networks with fixed and known node topology, in wireless networks it will not be known \emph{a priori} which user will connect to which BS, due to the undetermined user locations.

One way to deal with this problem is to employ a probabilistic caching policy~\cite{Blaszczyszyn2015optimal}, rather than the deterministic caching policy used in wired networks. To reflect the uncertainty, this approach treats the user locations as spatial random process, and then optimizes the probability of each content being cached at each BS.

A further complication arises when a user is moving from one cell to another during the duration of content delivery. If the user mobility pattern can be predicted, the caching decision can be optimized~\cite{poularakis2013exploiting}.
Otherwise, a less-than-ideal alternative is to model the user movement as Markov chain with  predictable probability transfer matrix.

\emph{Fading and interference:} In most of the current literature, the caching policy optimization fails to take into account channel fading and interference, which are essential to wireless networks. Consequently, the optimized policy may not perform well in practice. For example, distributed caching can increase cache-hit probability, which however may not improve SE and EE when path loss and ICI are considered~\cite{Dong}.\footnote{Cache-hit probability can reflect the cache utilization  efficiency, but does not necessarily reflect wireless resource utilization efficiency. This is an example of how content placement and content delivery are strongly entwined.} Specifically, when the nearest BS of a user does not cache the requested content of the user but the second nearest BS does, the signal power from the second nearest BS is lower due to path loss. Even worse, the nearest BS is possibly generating a strong interference towards the user.

\subsubsection{Caching at users}
The content placement at users in wireless networks differs from that in wired networks in both caching policy and the way to download contents. For precaching, the difference comes from fading and interference. For cache-enabled D2D, the difference comes from all mentioned aspects in wireless networks as in caching at the BSs.

\emph{Precaching:}
Traditional prefetching is implemented by over-the-top operators for the users who have installed applications (APPs)~\cite{Higgins12}.
Since the caching decision is made according to the predicted preference of a user and aims to improve QoE, the contents are naturally predownloaded to each user via unicast when the
channel is in good condition. This may generate interference to other on-going transmission.

When prefetching is implemented by a mobile operator who has knowledge of the congestion status of BSs and the channel conditions of users, the degradation of the network performance can be minimized by designing sophisticated scheduling.
Given that the predownloading via  unicast consumes the energy at both BS and user and may cause interference to other users, accurately predicting the contents that a user will
request and the time instance when the user will initiate  the request are important to guarantee that the advantages succeed the harms brought by prefetching.

Considering that content popularity usually changes slowly, the most popular contents can be pushed by broadcast at off-peak time, which causes negligible or even no performance
degradation to the network. For the contents (e.g., news) that update fast, a part of the network bandwidth can be reserved for
broadcasting to dynamically push the popular files to users~\cite{LHui14}; on the downside this may degrade the performance of
the network during peak time.

\emph{Cache-enabled D2D:}
Different from prefetching where the way to select contents for pre-downloading is ``selfish'', the principle of designing caching policy for D2D communications is to help other users.
Due to the user mobility, devices that will be in range of each other when a transmission request occurs are dynamic, which makes probabilistic
caching more appropriate. For a given popularity distribution,
the caching distribution that maximizes the offloading ratio takes on a form similar to \emph{water-filling}~\cite{JMY15}.

%
With the probabilistic caching policy, the contents can be downloaded to users by a combination of multicast and unicast. To saves wireless resources, the BS can transmit content to a user only when first requested, while the other users ``overhear" and cache content they find suitable; the gain of this approach is still unknown.

\section{Content Delivery: Impact of caching on wireless transmission}

To exploit the cache resource and traffic characteristics, some transmit strategies (essentially cross-layer) need to be re-designed.
Since caching at the BSs can replace backhaul and reduce latency while caching at the users can even offload wireless traffic,
new criteria emerge for optimization, e.g., minimizing backhaul traffic or end-to-end delay, and maximizing offloading ratio.

\subsection{User Association}
In heterogeneous networks with overlapped coverage of different BSs, caching may change the way of user association. To reduce end-to-end delay, or balance the traffic load in backhaul links, users are not always associated with the nearest BS. Instead, associating the users with the BSs that cache the requested contents may be more beneficial.
In this case, the closest BS to the user may generate interference stronger than the desired signal. Such a problem has been identified and studied in the literature.

\subsection{Coordinated Multi-Point Transmission (CoMP-JT) }
CoMP-JT is hard to implement in traditional networks due to the high capacity backhaul required for exchanging data among BSs.
Caching at SBSs makes CoMP-JT a possible cost-effective way to alleviate ICI, another limiting factor on wireless transmission capacity of small cell networks. If the requested contents for the users are cached at several adjacent BSs, the BSs can serve the users with CoMP-JT by only exchanging channel information~\cite{Lau13}. For helpers that completely lack backhaul, channel information needs to be shared with the assistance of the user device, or non-coherent CoMP-JT can be employed to avoid ICI and enhance the signal (but without any multiplexing gain).

Considering the limited cache size at SBSs and the dynamic nature of users' requests, cache-enabled CoMP-JT can only operate
opportunistically. To enjoy the gain from CoMP, the probability that the content
desired by a user can be found in local caches of multiple adjacent BSs should be increased. A simple way to do this is caching
popular contents in every SBS, which however inevitably reduces the overall cache-hit probability. To maximize SE of the cache-enabled CoMP-JT with given cache size, the caching policy needs to be carefully designed. Because not all users in each cell can be jointly served by adjacent SBSs with the requested contents locally cached, the SBSs not using CoMP-JT will generate interference to the CoMP users, which will limit the overall
throughput gain.

\subsection{Multicast}
Multicast is a mechanism to exploit content popularity for reducing duplicated transmission,
with which a BS can serve
multiple users requesting identical contents if they send the requests at the same time that happens for live video events.

The requests of users for content delivery are highly asynchronous.
For file downloading that is delay tolerant, this is not a big issue, where traditional multicast technology can be used to let the users wait for each other before starting transmission. For VoD streaming that has the particular QoE provision, this results in a large initial delay, which  degrades the user perceived QoE. Although schemes such as harmonic broadcasting partly overcomes this problem, they do not provide maximal SE.
This dilemma can be solved by jointly optimizing the content placement and content delivery. In~\cite{Ali13}, an ingenious coded-multicast strategy was proposed, which precaches partial contents possibly requested by all the users with a carefully designed network coding. Then, during the content delivery phase, different requests can be satisfied with a single multicast transmission. Despite the theoretical beauty and importance of this strategy, practical challenges remain, since the coding complexity grows exponentially and the download delay for each user increases with the number of users.

\subsection{Device-to-Device communications}
Traditional D2D communications can only offload peer-to-peer (P2P)
traffic between source and destination in proximity at the time they wish to communicate. Cache-enabled D2D creates new
opportunities to offload client/server traffic~\cite{Niki13}.

Nonetheless, different from wired P2P networks, in  D2D transmission there is not only a question of which users in vicinity are willing for help, but also the willingness of the users might change with time,
and specifically with the state of their battery. Consequently, incentivizing users to act as helpers is an important issue.
To justify the feasibility of cache-enabled
D2D communications, the energy consumed by D2D transmitters needs to be evaluated, and the caching policy needs to be jointly optimized with communication protocol to balance the throughput gain and the energy cost.
For readers' convenience, the key points of sections III and IV are summarized in Table \ref{tab:str}.

\section{Simulation Results}
To illustrate the design aspects in content placement and delivery, the performance difference between caching in wired and wireless edge, and quantitative results for some tradeoffs, we consider two representative systems. The simulation parameters are summarized in Table \ref{tab:simu}.

To compare the SE and EE gains of caching at wireless and wired edge over not caching, and show the SE/EE-cache size tradeoff, we consider a small cell network. Caches are either deployed at SBSs or the edge router as shown in Fig. \ref{fig:archi}. Each cache node caches the most popular files until it is full.

It is shown from Fig. 2 that both SE and EE benefit more from  caching at the wireless edge than at the wired edge. Specifically, the maximal SE gains are about $200\%$ and $900\%$ for wired and wireless edge caching, and the maximal EE gain are $200\%$ and  $500\%$, respectively. This is because compared with caching at the wired edge that can only relieve congestion in the mobile core network and Internet, caching at BSs can also relieve backhaul congestion, though the cache hit probability is lower. As the total cache size increases, SE increases for both wireless edge and wired edge caching. In both cases SE saturates, though the saturation point occurs at higher total cache size for wireless edge caching, since popular files need to be stored at multiple locations. EE generally also increases with total cache size, though EE first increases and then decreases for the case of very large library size (and thus large number of unpopular files), since the caches consume circuit power and caching rarely used files requires more energy than it saves ~\cite{Dong}.

To illustrate the design aspects except the cache size, we consider a cache-enabled D2D network, where each user sends $N_r$ requests sequentially in a period, and only the users within a  collaboration distance $r_c$ can establish D2D links. We also show the tradeoff between \emph{offloading ratio}, the amount of data conveyed by D2D divided by the total amount of data transmitted in the cell, and \emph{energy consumption}, which is the total transmit and circuit power consumed at each D2D transmitter averaged over channel fading, user location and file request. We consider \emph{optimal caching} proposed in~\cite{JMY15}, \emph{uniform caching} where each user caches files with identical probability, and \emph{popularity caching} where each user caches a file with probability equal to the file popularity.  We consider maximal transmit power (with legend ``$P_{\max}$'') and optimal transmit power (with legend ``$P^*$'') at each D2D transmitter, where $P^*$  is  obtained numerically by maximizing the offloading ratio.

It is shown from Fig. \ref{fig:d2d} that optimizing the caching policy can improve the offloading ratio and reduce the energy consumption, while optimizing the collaboration distance $r_c$ and transmit power are more critical. Furthermore, there is a tradeoff between the offloading ratio and the energy consumption when $r_c$ is small, as we mentioned in Section III-A-2.

\section{Conclusion Remarks}
Caching at the wireless edge can significantly improve SE and EE of wireless networks compared with caching at the wired edge. The gain comes from saving bandwidth and energy both for getting the files from the servers to the wireless infrastructure, as well as for the transmission from wireless infrastructure to users. To exploit the full potential of wireless edge caching, the unique limitations in wireless networks due to architecture and channel, such as topology, interference, users' mobility and limited battery must be considered for both content placement and delivery, and accurate predictions of popularity distributions and user preferences are critical.

\bibliographystyle{IEEEtran}
\bibliography{cachebib}

\begin{thebibliography}{10}
\providecommand{\url}[1]{#1}
\csname url@samestyle\endcsname
\providecommand{\newblock}{\relax}
\providecommand{\bibinfo}[2]{#2}
\providecommand{\BIBentrySTDinterwordspacing}{\spaceskip=0pt\relax}
\providecommand{\BIBentryALTinterwordstretchfactor}{4}
\providecommand{\BIBentryALTinterwordspacing}{\spaceskip=\fontdimen2\font plus
\BIBentryALTinterwordstretchfactor\fontdimen3\font minus
  \fontdimen4\font\relax}
\providecommand{\BIBforeignlanguage}[2]{{%
\expandafter\ifx\csname l@#1\endcsname\relax
\typeout{** WARNING: IEEEtran.bst: No hyphenation pattern has been}%
\typeout{** loaded for the language `#1'. Using the pattern for}%
\typeout{** the default language instead.}%
\else
\language=\csname l@#1\endcsname
\fi
#2}}
\providecommand{\BIBdecl}{\relax}
\BIBdecl

\bibitem{Niki13}
N.~Golrezaei, A.~F. Molisch, A.~G. Dimakis, and G.~Caire, ``Femtocaching and
  device-to-device collaboration: A new architecture for wireless video
  distribution,'' \emph{IEEE Commun. Mag.}, vol.~51, no.~4, pp. 142--149, Apr.
  2013.

\bibitem{Procach14}
E.~Bastug, M.~Bennis, and M.~Debbah, ``Living on the edge: The role of
  proactive caching in 5{G} wireless networks,'' \emph{{IEEE} Commun. Mag.},
  vol.~52, no.~8, pp. 82--89, Aug. 2014.

\bibitem{Ali13}
M.~A. Maddah-Ali and U.~Niesen, ``Fundamental limits of caching,'' \emph{IEEE
  Trans. Inf. Theory}, vol.~60, no.~5, pp. 2856--2867, May 2014.

\bibitem{JMY15}
M.~Ji, G.~Caire, and A.~Molisch, ``Wireless device-to-device caching networks:
  Basic principles and system performance,'' \emph{IEEE J. Sel. Areas Commun.},
  vol.~34, no.~1, pp. 176--189, Jan. 2016.

\bibitem{Dong}
D.~Liu and C.~Yang, ``Energy efficiency of downlink networks with caching at
  base stations,'' \emph{IEEE J. Sel. Areas Commun.}, vol.~34, no.~4, pp.
  907--922, Apr. 2016.

\bibitem{wang2014cache}
X.~Wang, M.~Chen, T.~Taleb, A.~Ksentini, and V.~Leung, ``Cache in the air:
  exploiting content caching and delivery techniques for {5G} systems,''
  \emph{IEEE Commun. Mag.}, vol.~52, no.~2, pp. 131--139, Feb. 2014.

\bibitem{Tatar14}
A.~Tatar, M.~D. de~Amorim, S.~Fdida, and P.~Antoniadis, ``A survey on
  predicting the popularity of web content,'' \emph{Springer Journal of
  Internet Services and Applications (JISA)}, vol.~5, no.~1, pp. 1--20, 2014.

\bibitem{Dey14}
H.~Ahlehagh and S.~Dey, ``Video-aware scheduling and caching in the radio
  access network,'' \emph{IEEE Trans. Netw.}, vol.~22, no.~5, pp. 1444--1462,
  Oct. 2014.

\bibitem{shi2014}
Y.~Shi, M.~Larson, and A.~Hanjalic, ``Collaborative filtering beyond the
  user-item matrix: A survey of the state of the art and future challenges,''
  \emph{ACM Comput. Surveys}, vol.~47, no.~1, pp. 3:1--3:45, May 2014.

\bibitem{Higgins12}
B.~D. Higgins, J.~Flinn, T.~J. Giuli, B.~Noble, C.~Peplin, and D.~Watson,
  ``Informed mobile prefetching,'' \emph{ACM MobiSys}, 2012.

\bibitem{LHui14}
K.~Wang, Z.~Chen, and H.~Liu, ``Push-based wireless converged networks for
  massive multimedia content delivery,'' \emph{IEEE Trans. Wireless Commun.},
  vol.~13, no.~5, pp. 2894--2905, May 2014.

\bibitem{Gharaibeh2015Provably}
A.~Gharaibeh, A.~Khreishah, B.~Ji, and M.~Ayyash, ``A provably efficient online
  collaborative caching algorithm for multicell-coordinated systems,''
  \emph{IEEE Trans. Mobile Computing}, to appear 2015.

\bibitem{Blaszczyszyn2015optimal}
B.~Blaszczyszyn and A.~Giovanidis, ``Optimal geographic caching in cellular
  networks,'' \emph{IEEE ICC}, 2015.

\bibitem{poularakis2013exploiting}
K.~Poularakis and L.~Tassiulas, ``Exploiting user mobility for wireless content
  delivery,'' \emph{IEEE ISIT}, 2013.

\bibitem{Lau13}
A.~Liu and V.~K. Lau, ``Mixed-time scale precoding and cache control in cached
  {MIMO} interference network,'' \emph{IEEE Trans. Signal Process.}, vol.~61,
  no.~24, pp. 6320--6332, Dec. 2013.

\end{thebibliography}

\newpage
\begin{IEEEbiography}[]
{Dong Liu} [S'13] (dliu@buaa.edu.cn) received the B.S. degree in electronics engineering from Beihang University (formerly Beijing University of Aeronautics and Astronautics), Beijing, China in 2013. He is currently pursuing Ph.D degree in signal and information processing with the School of Electronics and Information Engineering, Beihang University. His research interests lie in the area of caching in wireless network and cooperative communications.
\end{IEEEbiography}

\begin{IEEEbiography}[]
{Binqiang Chen} [S'14] (chenbq@buaa.edu.cn) received his B.S. degree in electronics engineering in 2012 and now is pursuing his Ph.D. degree in signal and information processing, both in the School of Electronics and Information Engineering, Beihang University, Beijing, China. His research interests include interference management, cooperative communications, device-to-device communications and content-centric networks.
\end{IEEEbiography}

\begin{IEEEbiography}[]
{Chenyang Yang} [M'99, SM'08] (cyyang@buaa.edu.cn) received the Ph.D. degree in electrical engineering from Beihang University (formerly Beijing University of Aeronautics and Astronautics), Beijing, China, in 1997. She has been a Full Professor with the School of Electronics and Information Engineering, Beihang University, since 1999. Her research interests include green radio, local caching, and other emerging techniques for next generation wireless networks.
She was the Chair of the IEEE Communications Society Beijing chapter from 2008 to 2012. She has served as a Technical Program Committee Member for numerous IEEE conferences. She has been an Associate Editor or a Guest Editor of several IEEE journals. She was nominated as an Outstanding Young Professor of Beijing in 1995 and was supported by the 1st Teaching and Research Award Program for Outstanding Young Teachers of Higher Education Institutions by Ministry of Education of China from 1999 to 2004.
\end{IEEEbiography}

\begin{IEEEbiography}[]
{Andreas F. Molisch} [S'89, M'95, SM'00, F'05] (molisch@usc.edu) is a Professor of Electrical Engineering at the University of Southern California. His current research interests are the measurement and modeling of mobile radio channels, ultrawideband communications and localization, cooperative communications, multiple-input-multiple-output systems, wireless systems for healthcare, and novel cellular architectures. He is a Fellow of NAI,  Fellow of AAAS, Fellow of IET, and Member of the Austrian Academy of Sciences, as well as recipient of numerous awards.
\end{IEEEbiography}

\newpage

\newpage
\begin{figure}[!htb]
	\centering
	\subfigure[SE-Cache size]{\includegraphics[width=0.75\textwidth]{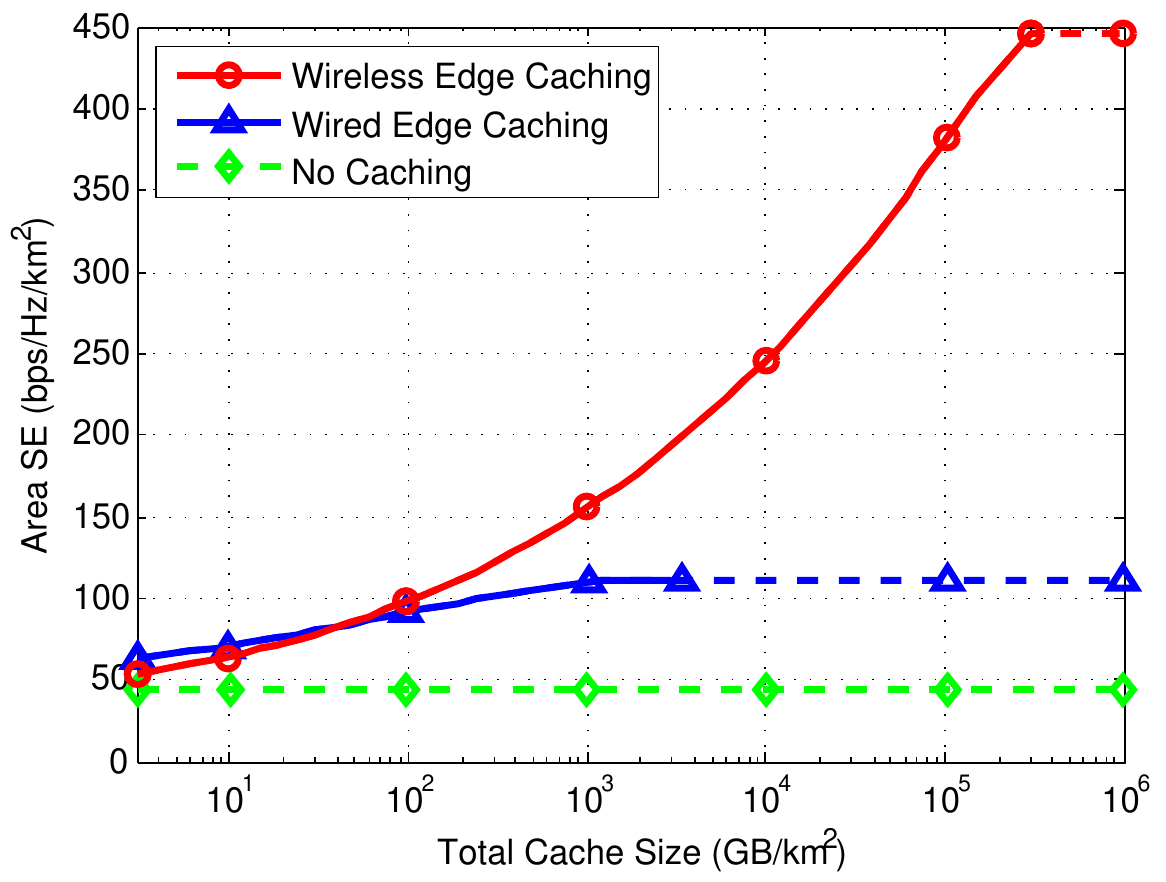} \label{fig:SE}}
	\subfigure[EE-Cache size]{\includegraphics[width=0.75\textwidth]{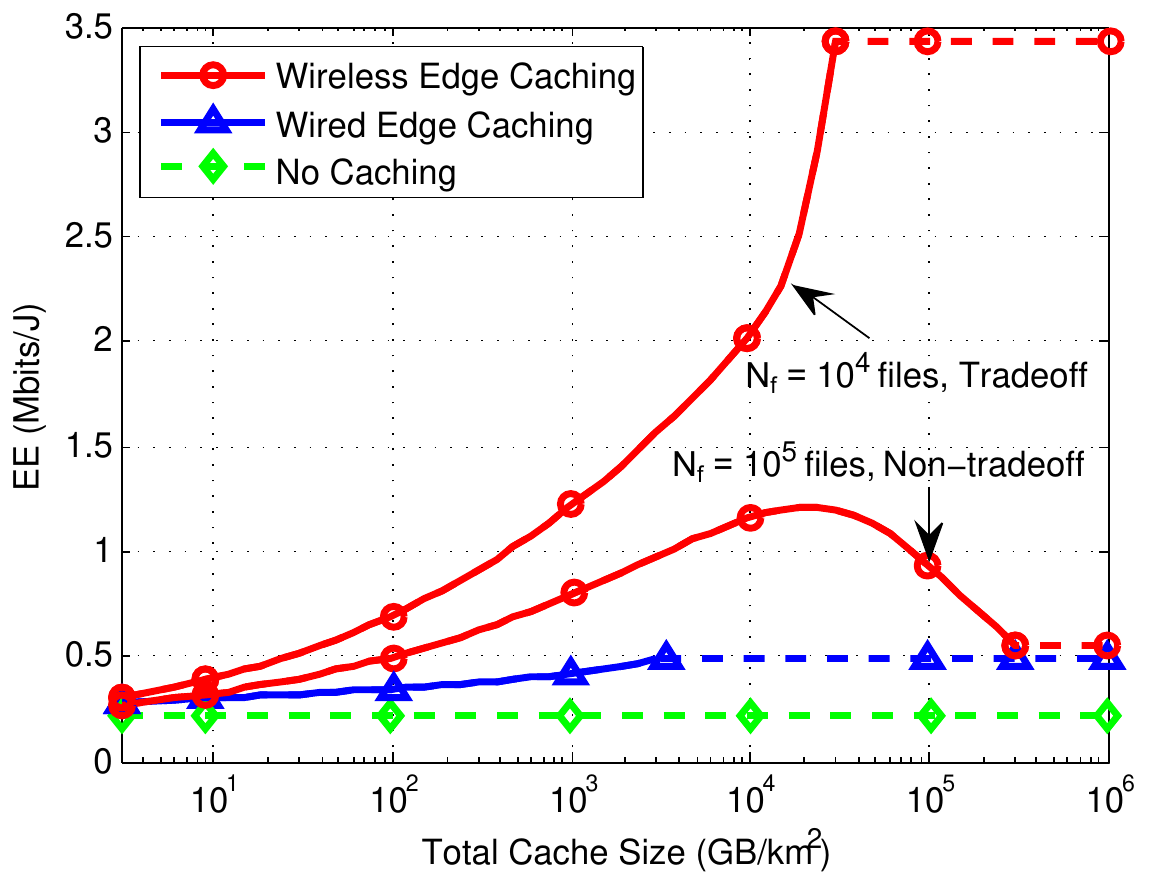} \label{fig:EE}}
	\caption{SE and EE comparison between caching at wireless and wired edge. The solid curves end when all the files have been cached at each cache node.}
	\label{fig:wireless} 
\end{figure}

\newpage
\begin{figure}[!htb]
	\centering
	\includegraphics[width=0.75\textwidth]{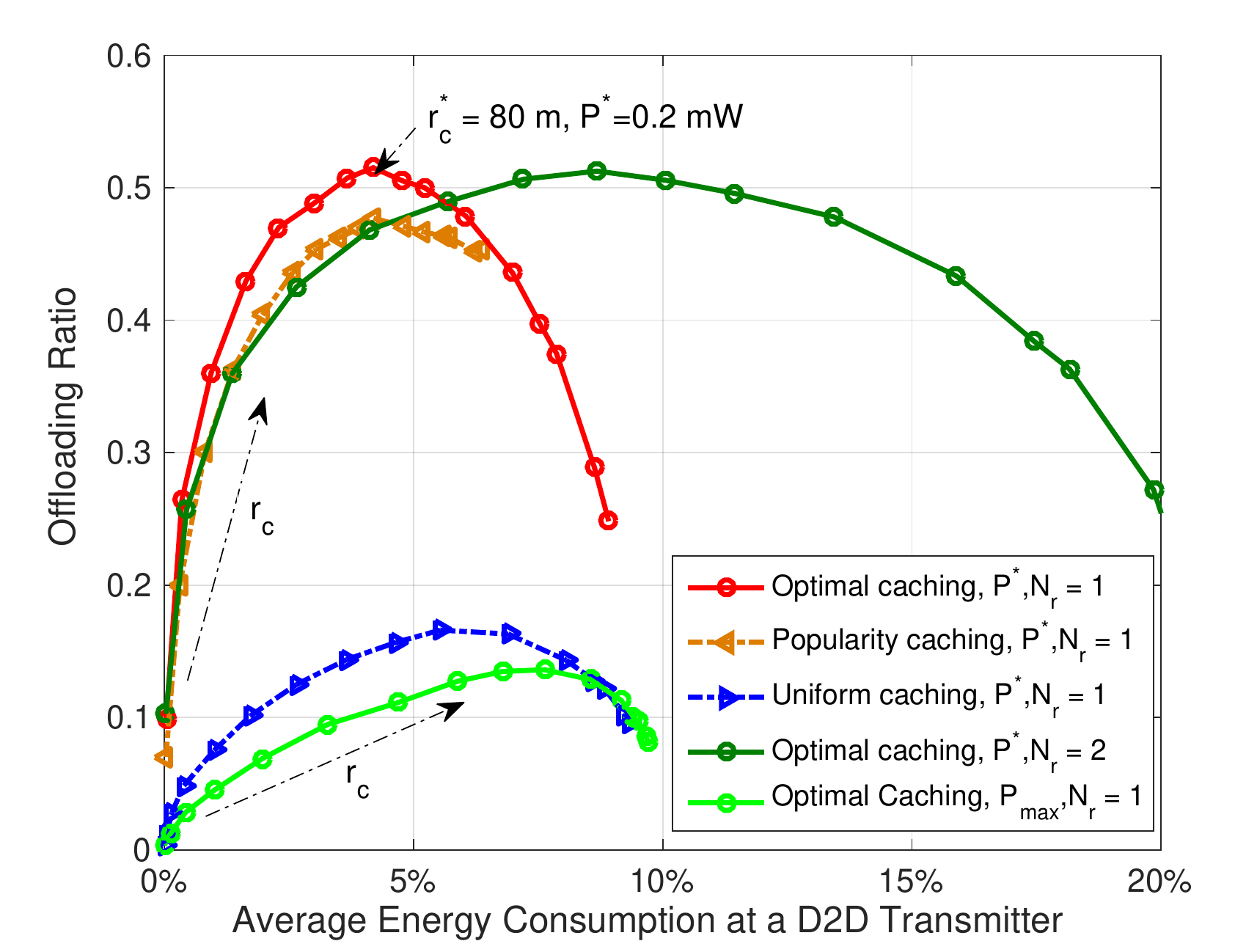}
	\caption{Offloading ratio and energy consumption. Each curve starts from $r_c = 10$ m and ends at $r_c = 400$ m, and $r_c^*$ is the optimal collaboration distance that maximizes the offloading ratio.}\label{fig:d2d}
\end{figure}

\newpage 
\begin{table}[ht]
	\centering
	\renewcommand{\arraystretch}{1.5}
	\caption{Tradeoffs, features and impacts of caching in wireless networks}
	\begin{tabular}{|m{2cm}|m{6cm}|l|l|}
		\hline
		\multirow{2}[4]{*}{} & \multirow{2}[3]{*}{\textbf{Caching at BSs}} & \multicolumn{2}{c|}{\textbf{Caching at users}} \\ \cline{3-4}
		&   & \textbf{Cache-enabled D2D} & \textbf{Precaching} \\\hline
		{\multirow{4}[-3]{2cm}{\textbf{Tradeoffs}}} & \multicolumn{3}{c|}{SE and cache size} \\ \cline{2-4}
		& EE and cache size & Offloading gain and energy cost & \multicolumn{1}{c|}{\multirow{2}[4]{*}{/}} \\ \cline{2-3}
		& Backhaul capacity and cache size & Throughput and outage & \\ \hline
		\multirow{5}[2]{2cm}{\textbf{Features in content placement}} & \multicolumn{2}{c|}{Small cache size and number of requests} & Small cache size \\ \cline{2-4}
		& \multicolumn{2}{c|}{Uncertainty in topology} &  \multicolumn{1}{c|}{/} \\ \cline{2-4}
		& \multicolumn{2}{c|}{Openness of wireless channel: a) overlapped coverage} & b) multicast/broadcast\\ \cline{2-4}
		& \multicolumn{3}{c|}{Fading and interference} \\ \cline{2-4}
		& \multicolumn{1}{c|}{/} & \multicolumn{2}{c|}{Limited battery capacity} \\ \hline
		\multirow{2}[3]{2cm}{\textbf{Impacts on wireless transmission}} & Optimization criteria: delay, backhaul traffic, offloading ratio & User incentive to act as helpers & \multicolumn{1}{c|}{\multirow{2}[3]{*}{/}} \\ \cline{2-3}
		& Transmission strategies: user association, CoMP-JT, multicast & Balancing offloading and energy &  \\
		\hline
	\end{tabular}%
	\label{tab:str}%
\end{table}%

\newpage
\begin{table*}[!htb]
	\centering
	\renewcommand{\arraystretch}{1.5}
	\caption{Simulation Parameters}
	\begin{tabular}{|l|l|l|}
		\hline
		\bf Systems &\bf Caching in SBSs  &\bf Cache-enabled D2D\\
		\hline
		File catalogue size, $N_f$ & $10^5$ & $10^3$ \\ \hline
		File size & \multicolumn{2}{c|}{30 MB} \\ \hline
		Zipf distribution skewness parameter, $\beta$ & \multicolumn{2}{c|}{0.8} \\ \hline
		Wireless transmission bandwidth & \multicolumn{2}{c|}{20 MHz (without interference coordination)} \\ \hline
		Transmit power of SBS or user & \multicolumn{2}{c|}{200 mW} \\ \hline
		Other power consumption parameters & See [5] and its reference [9] & Circuit power, 100 mW \\ \hline
		Considered region & 1 km$\times$1 km  &  0.5 km$\times$0.5 km   \\ \hline
		Number of SBSs (each with four antennas) & 100 (uniform-located)   & \multicolumn{1}{c|}{/} \\ \hline
		Number of single-antenna users & 300 (uniform-located)  & 2500 (uniform-located)\\ \hline
		Capacity of backhaul & 30 Mbps (microwave) & \multicolumn{1}{c|}{/}\\ \hline
		Transport bandwidth of wired network & 1 Gbps & \multicolumn{1}{c|}{/} \\ \hline
		Maximal energy allowed for transmitting one file  &\multicolumn{1}{c|}{/} & 10\% fraction of user battery capacity (1800 mAh) \\
		\hline
	\end{tabular}%
	\label{tab:simu}%
\end{table*}%
\end{document}